\documentclass[final,1p,times]{elsarticle} 

\usepackage{graphicx}
\usepackage{amssymb} 
\usepackage{amsthm} 
\usepackage{lineno}

\newcommand{\mnpart}{$\langle N_\mathrm{part} \rangle$}

\begin{document}

\begin{frontmatter} 

\title{Strange hadron and resonance production in Pb-Pb collisions at $\sqrt{s_{NN}}$ = 2.76 TeV with ALICE experiment at LHC}

\author{Subhash Singha (for the ALICE\fnref{col1} Collaboration)}
\address{Variable Energy Cyclotron Centre, 1/AF Bidhan Nagar, Kolkata - 700064, India}

\begin{abstract} 
  The ALICE experiment at the LHC has measured the production of strange hadrons and resonances in Pb-Pb and pp collisions at unprecedented high beam energies. The study of strange hadrons and resonances helps us to understand the properties of the medium created in the heavy-ion collisions and its evolution. We present the yields ($dN/dy$) at mid-rapidity  for strange hadrons ($\Lambda$, $\Xi^{-}$, $\Omega^{-}$,  their anti-particles and $K_{S}^{0}$ ) and resonances ($\phi$ and $K^{*0}$) for several collision centrality intervals. The results from Pb-Pb collisions at $\sqrt{s_{NN}}$ = 2.76 TeV are presented and compared to corresponding results from  pp collisions and lower energy measurements. Baryon to meson ratios and resonance to non-resonance particle ratios  relative to pp collisions are shown as a function of collision centrality and compared with the results at lower energies.
\end{abstract} 

\end{frontmatter} 

\section{Introduction}
The study of strange hadrons and resonances can provide information about the hot and dense strongly interacting matter created in high energy heavy-ion collisions. The colliding nuclei do not contain strange valence quarks. All the particles with non-zero strangeness quantum number are created in the course of the collision. The enhancement of strangeness in high energy heavy ion collisions relative to pp collisions is one of the signatures of quark gluon plasma formation \cite{qgp}. The measurement of resonance properties is also important because of their short lifetime (a few fm/$\it{c}$) and interactions with the medium. The hadronic decay products of the resonances can get rescattered in the hadronic medium causing a loss in the reconstructed resonance signal. Further a resonance may be generated due to the scattering among the hadrons in the medium. These two competing processes (rescattering and regeneration) play an important role in deciding the final resonance yields and hence the resonance to non-resonance particle ratios \cite{partratio}.
\section{Analysis Procedure}
The ALICE detector \cite{alicedet} at the LHC is designed to study both Pb-Pb and pp collisions at the TeV-scale centre of mass energy. The components of the ALICE detector used for the measurement of particle momenta and to reconstruct the primary vertex position are  the Inner Tracking System (ITS) and the Time Projection Chamber (TPC). The particle identification is done using the specific ionization energy loss inside the TPC.  A pair of scintillation hodoscopes, the VZERO detectors ( 2.8 $<$ $\eta$ $<$ 5.1 and -3.7 $<$ $\eta$ $<$ -1.7), were used for event triggering and centrality estimation. The strange hadrons and resonances are studied in a sample of approximately 10 million minimum bias $\sqrt{s_{NN}}$ = 2.76 TeV Pb-Pb events, collected during the year 2010.

                   The strange and multi-strange hadrons are measured through the reconstruction of the topology of their weak decays into charged particles only: $\Lambda \rightarrow p + \pi$, $K_{S}^{0} \rightarrow \pi^{+} + \pi^{-}$, $\Xi^{-} \rightarrow \pi^{-} + \Lambda$ and $\Omega^{-} \rightarrow K^{-} + \Lambda$ (and charge conjugates for anti-particle). The topological selection cuts are tuned in order to reduce the backgrounds without losing a considerable fraction of the signal.
The  resonances are reconstructed via their hadronic decay channels: $\phi \rightarrow K^{+} + K^{-}$ and  $K^{*0} \rightarrow K^{+} + \pi^{-}$ (and charge conjugate for anti-particle decays). The background is constructed using both the mixed event and like-sign techniques.
\begin{figure}[hbtp]
\begin{center}
\includegraphics[scale=0.097]{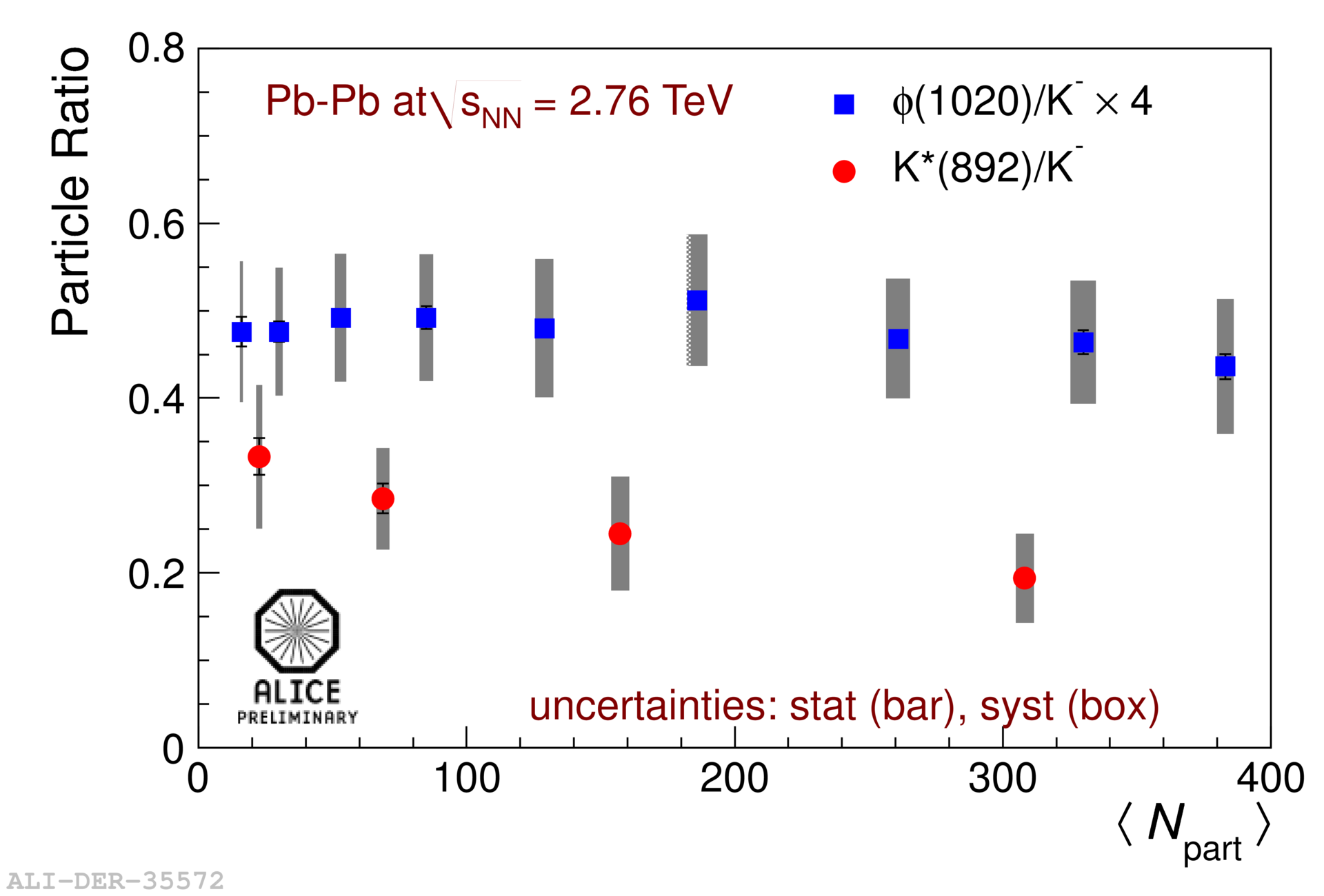}
\includegraphics[scale=0.094]{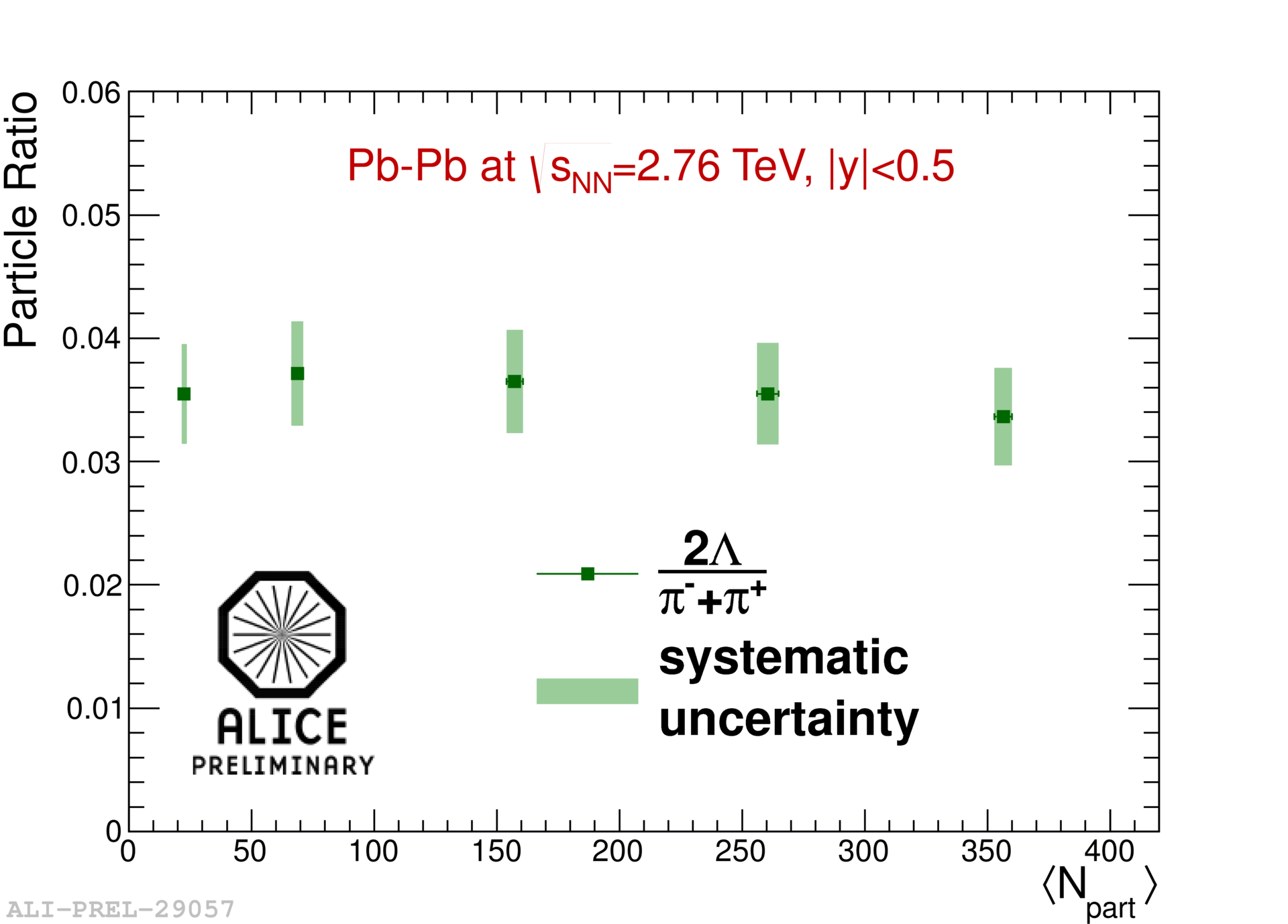}
\caption{(color online) Particle ratios as a function of mean number of participating nucleons (\mnpart) in Pb-Pb collisions at $\sqrt{s_{NN}}$ = 2.76 TeV. The left panel shows $\phi/K^{-}$ and $K^{*0}/K^{-}$ and the right panel shows $\Lambda/\pi$ as a function of \mnpart. The bands denote the systematic uncertainties (uncorrelated bin to bin) and the bars denote the statistical uncertainties.}
\end{center}
\end{figure}
\section{Results and discussions}
The transverse momentum spectra are obtained for strange hadrons ($\Lambda$, $\Xi^{-}$, $\Omega^{-}$,  their anti-particles and $K_{S}^{0}$ ) and resonances ($\phi$ and $K^{*0}$) in different collision centralities in Pb-Pb collisions at 2.76 TeV after correcting for reconstruction efficiency, detector acceptance and branching ratio. In order to extract the particle yields of strange hadrons and resonances, the spectra are fitted using a Blast-Wave parameterization \cite{bgbw} (Tsallis-Levy parameterization \cite{tsallis} is used for $K^{*0}$){\footnote[1]{Since the $K^{*0}$ spectra is not well described by the Blast-Wave parameterization for the peripheral collisions.}}. The yields are calculated integrating the corrected spectra in the measured $p_{T}$ region and extrapolating using the fit outside this region. To evaluate the systematic uncertainty in the extrapolation region, exponential and Tsallis-Levy functions were used. 
\begin{figure}[ht]
\begin{center}
\includegraphics[scale=0.099]{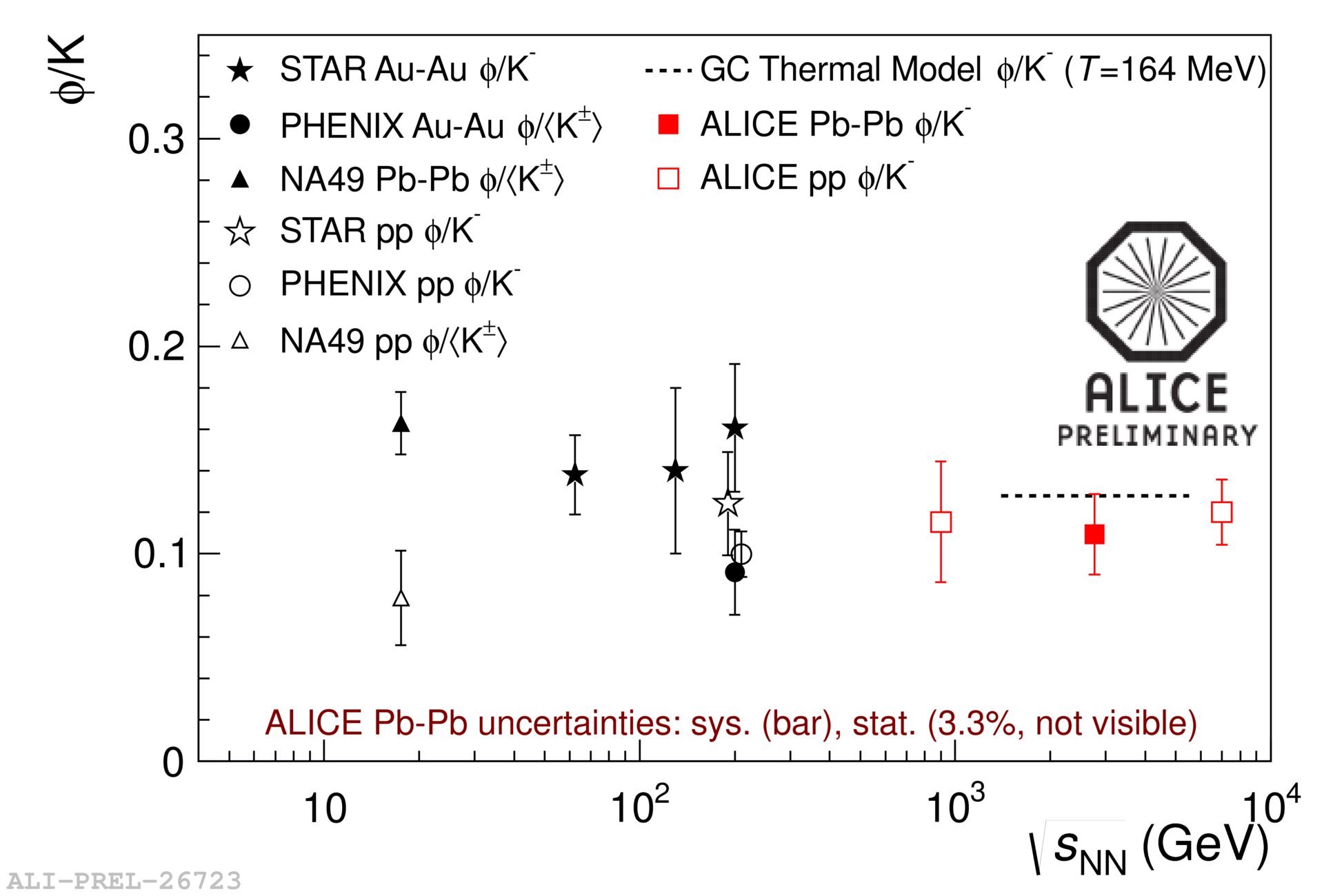}
\includegraphics[scale=0.099]{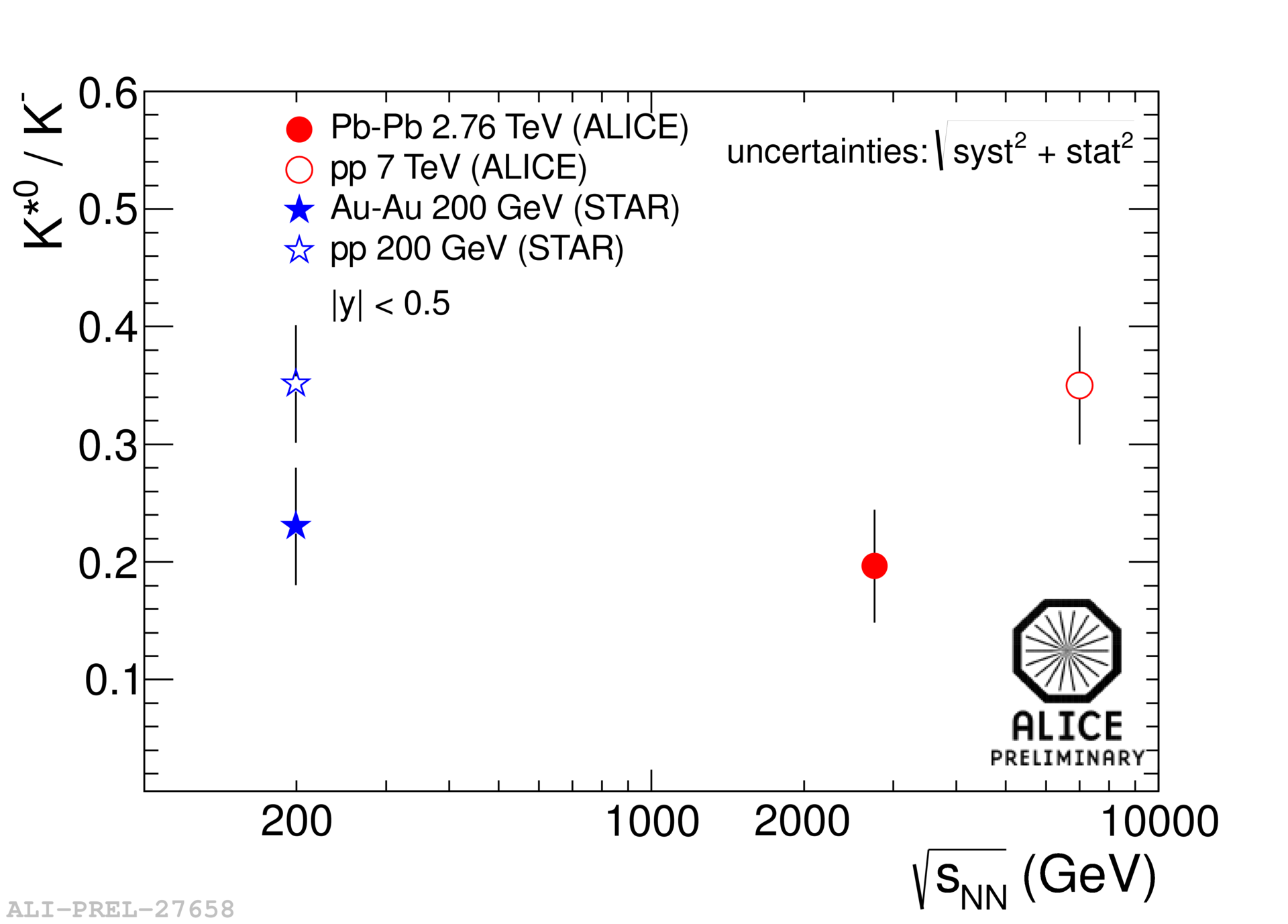}
\includegraphics[scale=0.099]{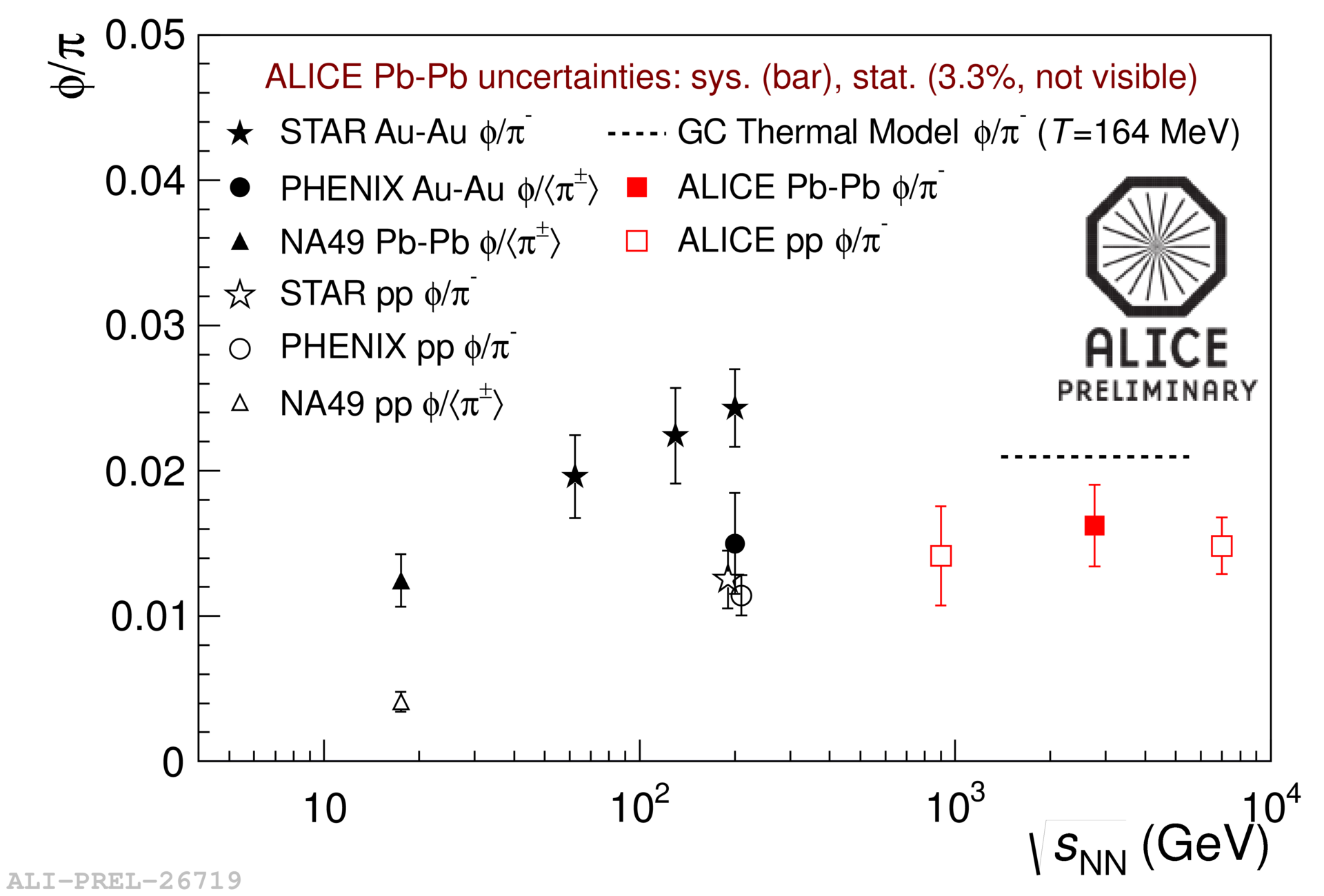}
\includegraphics[scale=0.099]{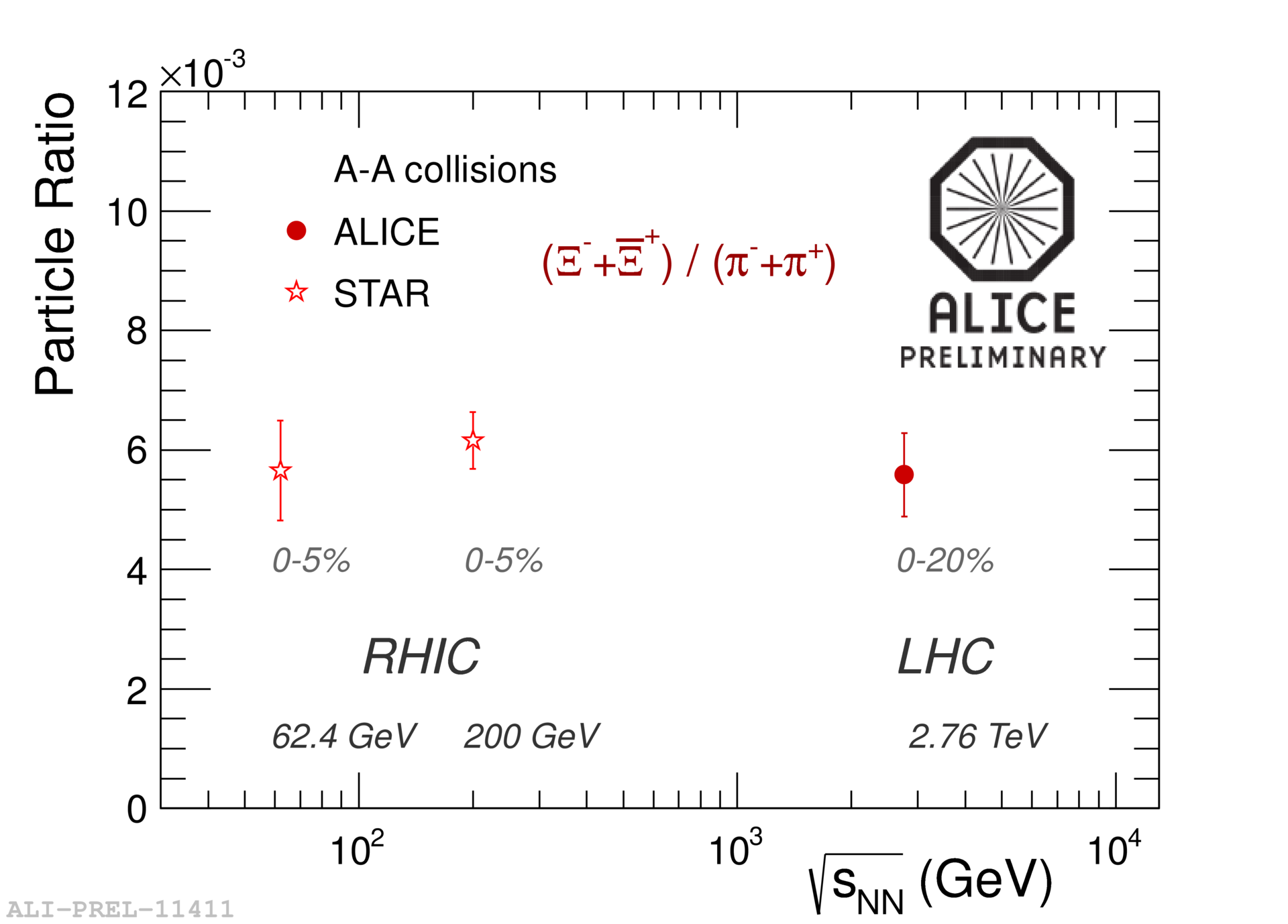}
\caption{(color online) Particle ratios $\phi/K$, $K^{*0}/K^{-}$, $\phi/\pi$ and $\Xi/\pi$ as a function of beam energy. The solid red points are the measurements in Pb-Pb collisions in ALICE. The pp measurements in ALICE are shown by the open red squares in the top left and the bottom left panels and the open red circle in the top right panel.}
\end{center}
\end{figure}
Figure 1 shows the particle yield ratios $\phi/K^{-}$, $K^{*0}/K^{-}$ and $\Lambda/\pi$ as a function of mean number of participating nucleons (\mnpart) in Pb-Pb collisions at $\sqrt{s_{NN}}$ = 2.76 TeV. A weak centrality dependence is observed in the $K^{*0}/K^{-}$ ratio, while the $\phi/K^{-}$ ratio is independent of collision centrality. The decreasing trend in $K^{*0}/K^{-}$ ratio suggests (considering the factor of 10 difference in lifetime of $\phi$ and $K^{*0}$)  a possible increase in rescattering in the most central collisions. A naive expectation from a kaon coalescence model \cite{kaoncoal} is an increase in $\phi/K$ ratio with increasing collision centrality, which is not seen in our data. The $\Lambda/\pi$ ratio is found to be independent of collision centrality. We have also studied the beam energy dependence of particle yield ratios in Figure 2. The results are compared with lower energy measurements at SPS \cite{spspp} and RHIC \cite{rhicpp} and also with the pp measurements \cite{ppalice}. The particle ratios $\phi/K$, $\phi/\pi$, $K^{*0}/K^{-}$ and $\Xi/\pi$ are found to be independent of beam energy. The $K^{*0}/K^{-}$ ratio in Pb-Pb collisions is less than that measured in pp collisions. This observation may also indicate the effect of hadronic rescattering at LHC energies.
\begin{figure}[hbtp]
\begin{center}
\includegraphics[scale=0.13]{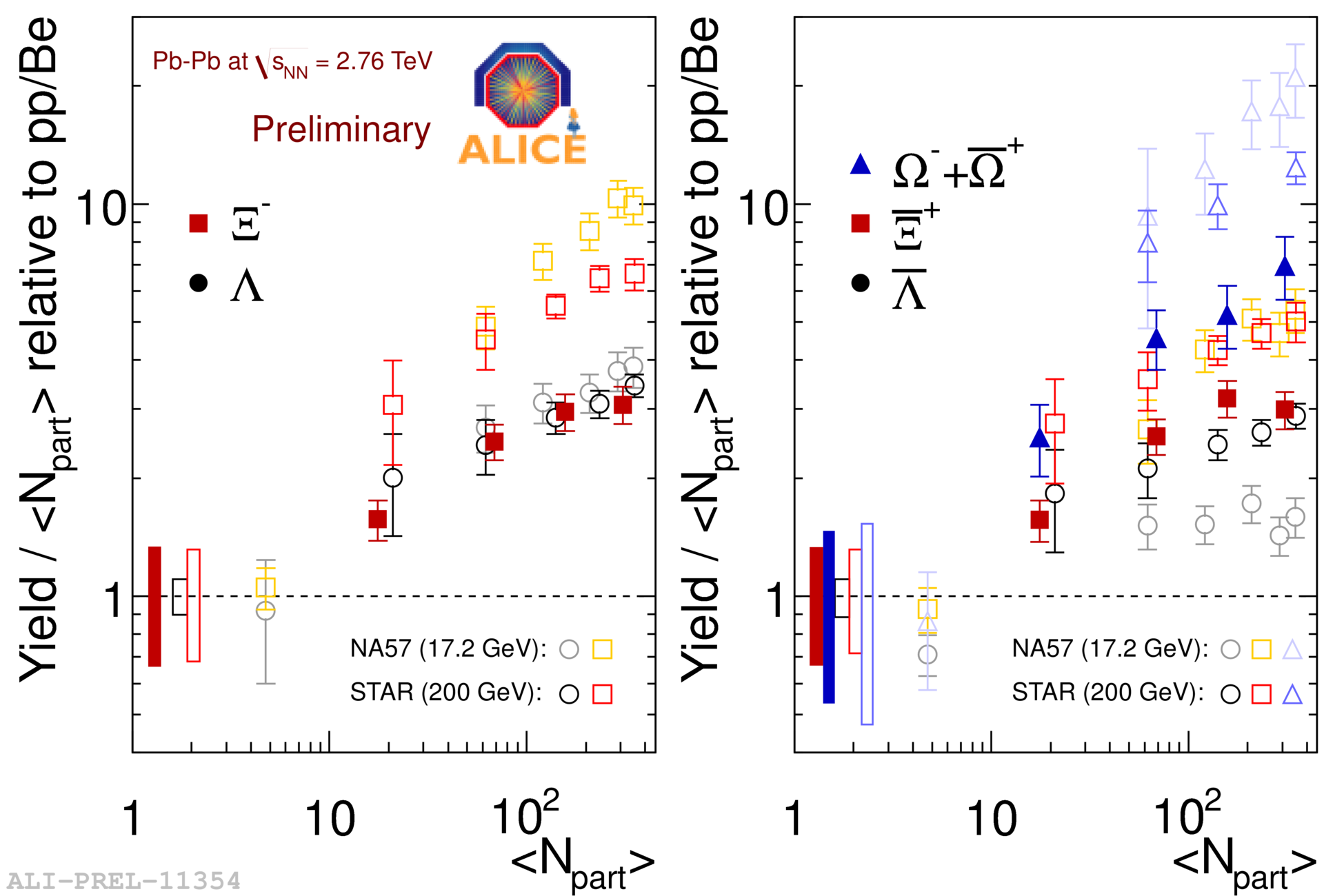}
\caption{(color online) Strangeness enhancement as a function of mean number of participating nucleons (\mnpart) in Pb-Pb collisions at $\sqrt{s_{NN}}$ = 2.76 TeV. The results are compared with measurements at SPS and at RHIC (open symbols). The bars indicate the systematic uncertainty on the pp (pBe for the SPS data) reference. (pp reference values obtained by interpolating the ALICE data at the two energies 0.9 and 7 TeV \cite{strangepp} for $\Xi$, while the STAR data at 200 GeV \cite{rhic} and the ALICE data at 7 TeV for $\Omega$)}
\end{center}
\end{figure}

        The strangeness enhancement has been calculated as the yield per participant in Pb-Pb collisions relative to the yield per participant in pp collisions. Figure 3 shows the enhancement of $\Xi^{-}$, $\Omega^{-}$ as a function of \mnpart. These results are compared with the measurements at SPS \cite{sps} and at RHIC \cite{rhic}. We observe that the enhancement in $\Omega^{-}$ is larger than that of $\Xi^{-}$,  following the hierarchy based on the strangeness content of the particle. The enhancement decreases with the increase in the beam energy following the trend already observed between SPS and RHIC.
\section{Conclusions}
The invariant yields at mid-rapidity have been measured in Pb-Pb collisions at 2.76 TeV for strange hadrons and resonances. No centrality dependence is observed in $\phi/K^{-}$ and $\Lambda/\pi$ ratios, whereas a weak centrality dependence is observed in $K^{*0}/K^{-}$ which suggests a possible increase in rescattering for central collisions. No energy dependence is observed in the particle ratios $\Xi/\pi$, $\phi/\pi$, $\phi/K$ and $K^{*0}/K^{-}$. The $K^{*0}/K^{-}$ ratio in heavy ion collisions is found to be less than that measured in pp collisions: this observation may again point to the effects of hadronic rescattering at LHC energies. Strangeness enhancement is studied with the multi-strange baryons ($\Xi$ and $\Omega$) and compared to lower energy measurements. The enhancement decreases with the increase in the beam energy and it is found to increase with the increase in the strangeness content.






\begin{thebibliography}{20} 

\bibitem{qgp} J. Rafelski, B. Muller {\it{Phys. Rev. Lett.}} 48, 1066(1982); P. Koch, B. Muller, J. Rafelski {\it{Phys. Rep.}} 142, 167 (1986).
\bibitem{partratio} G. Torrieri and J. Rafelski {\it{Phys. Lett.}} B 509, 239-245 (2001).
\bibitem{alicedet} K. Aamodt {\it{et. al.}} [ALICE collaboration], {\it{J. Inst. }} 3 No. SO8008 i-245.
\bibitem{zdc} K. Aamodt {\it{et. al.}} [ALICE collaboration], {\it{Phys. Rev. Lett.}}  105, 032301 (2011).
\bibitem{bgbw} E. Schnedermann, J Sollfrank, and U. Heinz {\it{Phys. Rev. C}}  48, 2462-2475 (1993).
\bibitem{tsallis} C. Tsallis {\it{J. Stat. Phys}}  42, 479 (1988).
\bibitem{starphik} M. M. Aggarwal {\it{et. al.}} [STAR collaboration], {\it{Phys. Rev. C}}  84, 034909 (2011); J Adams {\it{et. al}} [STAR collaboration] {\it{Phys Lett B}} 612, 181(2005).
\bibitem{kaoncoal} H. Sorge {\it{ Phys Rev }} C 52, 3291(1995); M. Bleicher {\it{et. al.}} {\it{J. Phys G}} 25, 1859(1999); Y. Lu {\it{et. al.}} {\it{J. Phys G}} 32, 1121(2006); B. Mohanty and N. Xu {\it{J. Phys G}} 36, 064022(2009).
\bibitem{spspp} S. V. Afanasiev {\it{et. al.}} [NA49 collaboration], {\it{Phys. Lett.}} B 491, 59-66(2000).
\bibitem{rhicpp} B. I. Abelev {\it{et. al.}} [STAR collaboration], {\it{Phys. Rev.}} C 79, 64093(2011); A. Adare {\it{et. al.}} [PHENIX collaboration], {\it{Phys. Rev.}} C 83, 64093(2011).
\bibitem{ppalice} B. Abelev {\it{et. al.}} [ALICE collaboration], arXiv: 1208.5717v1[hep-ex] (2012).
\bibitem{strangepp} B. Abelev {\it{et. al.}} [ALICE collaboration], {\it{Phys. Lett.}} B 712, 309-318(2012).
\bibitem{sps} E. Anderson {\it{et. al.}} [WA97 collaboration], {\it{Phys. Lett.}} B 449, 401(1999); F. Antinori {\it{et. al.}} [WA97/NA57 collaboration], {\it{Nucl Phys}} A 698, 118(2002).
\bibitem{rhic} B. Abelev {\it{et. al.}} [STAR collaboration], {\it{Phys. Rev.}} C 77, 044908 (2008).
\end{thebibliography}
\end{document}